\documentclass[preprint,12pt]{elsarticle}

\newcommand{\ga}{\gamma}

\newcommand{\om}{\omega}

\newcommand{\prt}{\partial}

\usepackage{amssymb}
\usepackage{amsmath}

\begin{document}

\title{Solitons in a one-dimensional rhombic waveguide array}

\author{D.~V.~Shaykin}
\affiliation{Russian University of Transport (RUT-MIIT), Obrazcova st. 9, Moscow, Russia}
\affiliation{Skolkovo Institute of Science and Technology, Skolkovo, Moscow, 143026, Russia}
\affiliation{Moscow Institute of Physics and Technology, Institutsky lane 9, Dolgoprudny, Moscow region, 141700, Russia}

\begin{abstract}
Two types of soliton solutions are analytically considered in a rhombic one-dimensional lattice: transverse (discrete) solitons and longitudinal solitons. Based on the multi-scale method, longitudinal solitons are obtained as envelopes of wave packets outside the forbidden gap, on the gap and under the gap. A discrete soliton was obtained based on a wave packet from the center of the Brillouin zone. The numerical calculations are in good agreement with the analytical predictions. 
\end{abstract}


\maketitle

\section{Introduction}
 To study various physical processes, a popular method is to create their simulation (e.g., \cite{1,2,3,4}). Thus, a convenient physical basis for simulating various discrete phenomena is the consideration of an array of waveguides (e.g., \cite{8,11,12}).
Initially, the research activity on the behavior of light in waveguides shifted from ordinary chains \cite{15,16} of waveguides to chains with more interesting properties\cite{19,20} and complex geometries \cite{17,18}, including rhombic ones \cite{21,22,23,24}.

\begin{figure}
    \centering
    \includegraphics[scale = 0.25]{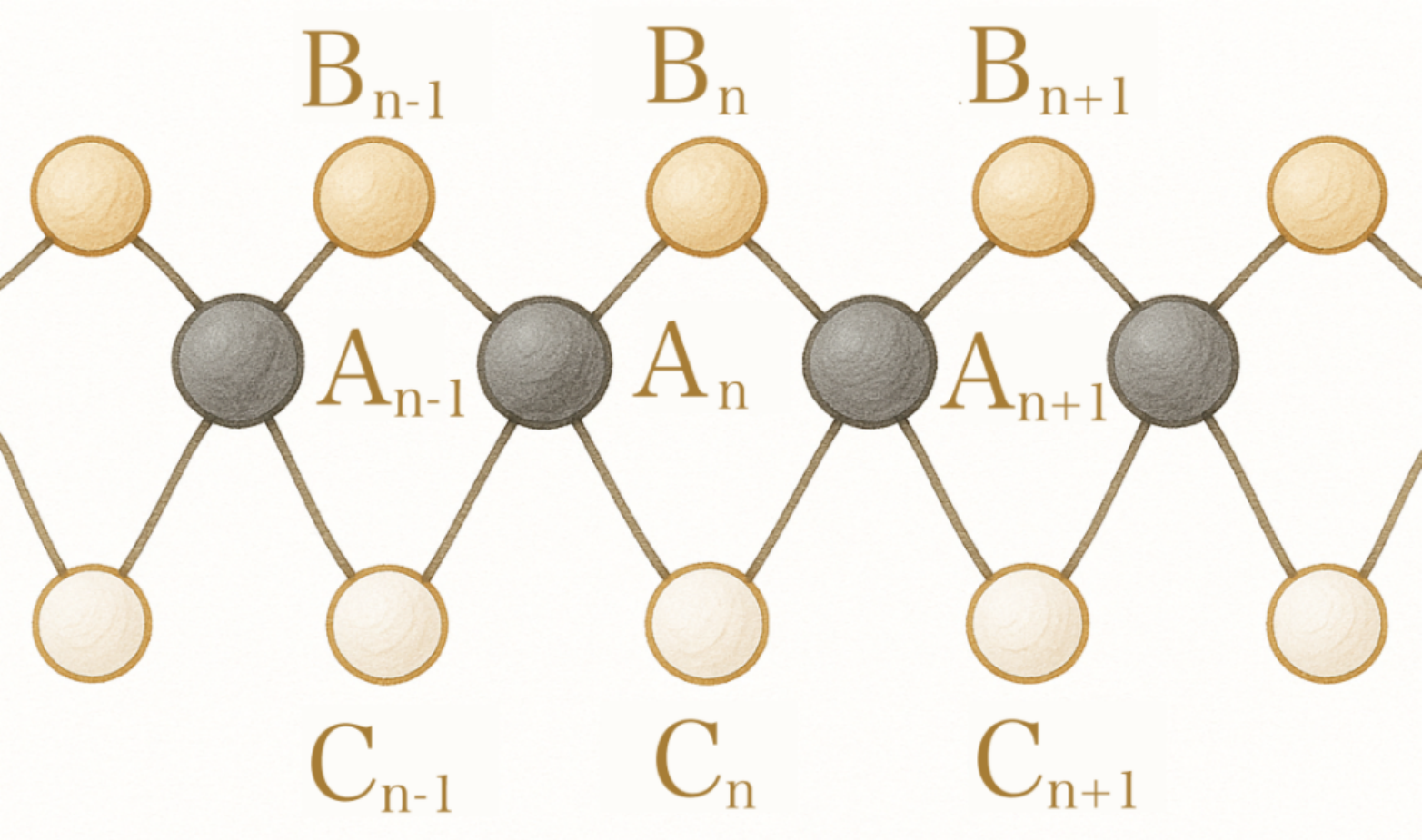}
    \caption{Rhombic matrix of waveguides with different coupling coefficients $A-B$ and $A-C$.}
    \label{fig0}
\end{figure}

Here we will present an analytical derivation for the various types of solitons characteristic of the rhombic array of waveguides with opposite refractive indices described by the system: 
\begin{equation}\label{base1}
\begin{split}
&i\frac{dA_n}{dz} + \big( B_n+B_{n-1}\big)+\ga\big( C_n + C_{n-1} \big)+
\mu_a |A|^2A = 0, \\
&-i\frac{dB_n}{dz}+\big( A_n + A_{n+1} \big)+ \mu_b |B|^2B = 0,\\
&-i\frac{dC_n}{dz}+\ga\big( A_n + A_{n+1} \big)+ \mu_c |B|^2B = 0
\end{split}
\end{equation}
where $\pm d/dz = \prt/\prt t \pm \prt/\prt x$ and the corresponding picture of waveguides distribution is shown in Figure \ref{fig0}. A dispersion relation for linear waves $\propto e^{i(kx-\om t + qn)}$ has two branches: $\om = -k$ and
\begin{equation}\label{disp1}
    \om^2 = k^2 + |1+e^{iq}|^2(1+\gamma^2).
\end{equation}
Previously, a soliton solution was obtained \cite{26} only in the case of $\mu_b=\mu_c =0$, in which the system \eqref{base1} was reduced to the case of two basic waveguides.

\section{Solitons along the fibers}
Let's move from the array of waveguides \eqref{base1} to the basic cell by setting 
\begin{align}
A_n = Ae^{iqn}, \quad B_n = Be^{iqn}, \quad C_n = Ce^{iqn}.
\end{align}
This leads to three equations, the solutions of which we will look for in the form of wave packets \cite{27,28}:
\begin{align}
&A(x,t) = \tilde{A}(\xi)e^{i(kx-\om t +\phi(\xi))}, \\
&B(x,t) = \big[ \tilde{B}(\xi)+i\tilde{b}(\xi)\big]e^{i(kx-\om t + \phi(\xi))},\\
&C(x,t) = \big[ \tilde{C}(\xi)+i\tilde{c}(\xi)\big]e^{i(kx-\om t + \phi(\xi))},
\end{align}
where $\xi = x-Vt$. After substitution, we get
\begin{multline}\label{system1}
   i(1-V)\dot{\tilde{A}}-\big[ k-\om + \dot{\phi}(1-V) \big]\tilde{A}+(1+e^{iq})\big[\tilde{B}+i\tilde{b}+\gamma(\tilde{C}+i\tilde{c})\big]+\mu_a\tilde{A}^3=0,
    \\   
   -i(1+V)\big[ \dot{\tilde{B}}+i\dot{\tilde{b}} \big] + (k+\om+\dot{\phi(1+V)})\big[\tilde{B}+i\tilde{b}\big]+
    \tilde{A}(1+e^{iq})+\mu_b  \big[\tilde{B}^2+\tilde{b}^2\big](\tilde{B}+i\tilde{b})=0,
    \\
   -i(1+V)\big[ \dot{\tilde{C}}+i\dot{\tilde{C}} \big] + (k+\om+\dot{\phi(1+V)})\big[\tilde{C}+i\tilde{c}\big]+
  \tilde{A}(1+e^{iq})+\mu_c  \big[\tilde{C}^2+\tilde{c}^2\big](\tilde{C}+i\tilde{c})=0.
    \\
\end{multline}

Separating the imaginary and real contributions, we obtain six equations, which we will work with in this section:
Here it will be convenient for us to use the dispersion law \eqref{disp1} in the form:
\begin{equation}\label{disp2}
\om^2 = k^2+(\alpha^2+\beta^2)(1+\gamma^2);
\end{equation}
where $\alpha  = 1+\cos{q}$, $\beta = \sin{q}$.

To begin with, we study wave packets with exponential amplitude modulation. To do this, we assume the amplitude functions are proportional to $\exp{(\Omega \xi)}$ and linearize the system of six equations obtained from Eqs.\eqref{system1}. This linear system does not contain $\phi$, since it was included in the form of multiplication by amplitude functions. Thus, we have five unknowns and six equations in a system whose compatibility is possible only under the fulfillment of two conditions:
\begin{equation}
k^2 = \frac{\Omega^2+\om^2-(\alpha^2+\beta^2)(1+\gamma^2)}{1+\frac{\Omega^2}{\om^2}},
\end{equation}
\begin{equation}\label{Vel}
V = k/\om.
\end{equation}
 Their meaning is easily revealed in the limit of $\Omega \to 0$: the first is the dispersion law written in an inverted form, and the second is the group velocity. This means that we can have wave packets with a carrier wavenumber inside the band gap if the amplitude is modulated exponentially, that is, in a typical situation for a soliton.

In addition, it is natural to assume that $\Omega = \varepsilon\om$, $\varepsilon \ll1$ (since we are considering wave packets),  and to study the envelope, we should switch to the slow variable $\Omega\xi$.

\subsection{ Outside the gap ($\omega>\delta$).}
Since there is a small parameter $\varepsilon$ in our problem, we will look for a solution in the form of series
\begin{equation}
\{\tilde{A},\tilde{B},\tilde{b},\tilde{C},\tilde{c},\phi\} = \sum_{n=1}^{\infty}\varepsilon^n\{\tilde{A}_n,\tilde{B}_n,\tilde{b}_n,\tilde{C}_n,\tilde{c}_n,\phi_n\} ,
\end{equation} 
We also need to decompose the dispersion law:
\begin{equation}
k = \sqrt{\om^2-\delta^2}+\frac{\delta^2}{2\sqrt{\om^2-\delta^2}}\varepsilon^2+o(\varepsilon^2)
\end{equation}
where for brevity we introduce $(\alpha^2+\beta^2)(1+\gamma^2) =\delta^2$, $\sqrt{\om^2-\delta^2} = k_0$, $\delta^2/2\sqrt{\om^2-\delta^2} = k_2$.

Substituting these expansions into system \eqref{system1} and separating the imaginary and real parts, as well as separately considering the expressions for each degree $\varepsilon$, we obtain the following results:

$\varepsilon:$
\begin{equation}\label{1e1}
\tilde{B}_1=\tilde{C}_1/\gamma= -\frac{\alpha}{\omega+k_0}\tilde{A}_1, \quad \{\tilde{b}_1,\tilde{c}_1\} = \frac{\beta}{\alpha}\{\tilde{B}_1,\tilde{C}_1\}
\end{equation}

$\varepsilon^3:$
\begin{equation}
\ddot{\tilde{A}}_1 = \tilde{A}_1-2\theta (\tilde{A}_1)^3
\end{equation}
It is a well-known oscillation equation with the soliton solution, where we put
\begin{equation}\label{T1}
\theta = \frac{\delta^2}{2(1+\gamma^2)^2(\om+k_0)^3}\bigg[\mu_b+\gamma^4 \mu_c+\mu_a\bigg(\frac{(\om + k_0)\sqrt{1+\gamma^2}}{\delta}\bigg)^2
\bigg]
\end{equation}
Thus, in the first order of $\varepsilon$, we obtain
\begin{equation}\label{1e3}
\tilde{A} = \frac{\varepsilon e^{i(kx-\om t)}}{\sqrt{\theta}\cosh{(\Omega \xi/2)}},
\end{equation}
and Eqs.\eqref{1e1} allow us to obtain the solitons in other optical fibers.

The number of oscillations inside the soliton depends on a small parameter such as $\Omega \sim 1/\varepsilon$ and the amplitude such as $\sim \varepsilon$.

\subsection{On the gap, inside the gap ($\om\leq\delta$)}

Decomposition of the law of dispersion on the gap is
\begin{equation}
k = \varepsilon\delta+ o(\varepsilon^2) = \varepsilon k_1+o(\varepsilon^2),
\end{equation}
where $k_1 = \om = \delta$.

Decomposition inside the gap is possible if we put $\om = \delta-d\varepsilon^2$:
\begin{equation}
k = \varepsilon \sqrt{\delta^2-2d\delta}+ o(\varepsilon^2) = \varepsilon k_1+o(\varepsilon^2).
\end{equation} 
where $k_1 = \sqrt{\delta^2-2d\delta}$, which means $d\lesssim \delta/2$.
Since the decompositions are identical up to the notation, the system of equations \eqref{system1} after substitution for these two cases must match:

$\varepsilon:$
\begin{equation}
\tilde{B}_1=\tilde{C}_1/\gamma = -\frac{\alpha}{\delta}\tilde{A}_1, \quad \{\tilde{b}_1,\tilde{c}_1\} = \frac{\beta}{\alpha}\{\tilde{B}_1,\tilde{C}_1\}
\end{equation}

$\varepsilon^3:$
\begin{equation}
\ddot{\tilde{A}}_1 = \tilde{A}_1-2\theta (\tilde{A}_1)^3,
\end{equation}
where
\begin{equation}\label{T2}
\theta = \frac{1}{\delta(1+\gamma^2)^2}\bigg[ (1+\gamma^2)^2\mu_a + \mu_b+\gamma^4 \mu_c\bigg].
\end{equation}
The type of solution is the same as that described in equation \eqref{1e3}. The difference is that the characteristic number of oscillations carried by this soliton is on the order of $k/\Omega \sim 1$. If we put $d = \delta/m$ for simplicity, then the velocity of the soliton is approximately equal to $\varepsilon \sqrt{1-2/m} $ according to Eq. \eqref{Vel}.

\begin{figure}
\begin{center}
\begin{picture}(130.,120.)
\put(-120,0){\includegraphics[width = 6.15cm]{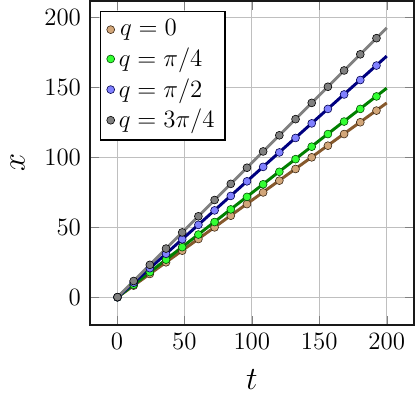}}
\put(70,0){\includegraphics[width = 6cm]{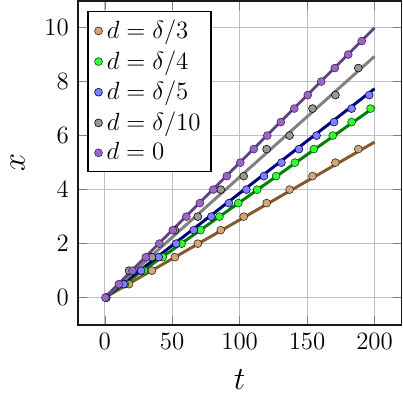}}
\put(-16,182){(a)}
\put(163,182){(b)}
\end{picture}
\caption{
Graphs of the dependence of the trajectories of solitons ((a) - outside the gap, (b) - inside the gap); circles indicate numerical data, and solid lines indicate an analytical result, according to formula \eqref{Vel}. These data correspond to $\gamma = 1.5, \varepsilon = 0.05$ and for (a) we put $\om = 5$, for (b) we put $q=0$.}
\label{fig12}
\end{center}
\end{figure}

\section{Discrete solitons}
Now we will carry out similar arguments for discrete solitons. First, we will find the law of dispersion of waves slightly modulated exponentially in amplitude as $e^{i(kx-\om t +qn)-\varepsilon n}$: 
\begin{equation}
    k = \sqrt{\om^2-2(1+\gamma^2)-2(1+\gamma^2)\cos{(q+i\varepsilon)}}
\end{equation}
It's easy to see that non-damping solutions can only occur in the case of $q=0,\pi$. Next, we will only work with the case $q=0$ with decomposition
\begin{equation}
    k = k_0-k_2\varepsilon^2+o(\varepsilon^2),
\end{equation}
$k_0 = \sqrt{\om^2-4(1+\gamma^2)}$, $k_2 = (1+\gamma^2)/2k_0$, as $q=\pi$ does not lead to the desired solution.

We will look for stationary solitons with an amplitude that depends only on the waveguide number as $A_ne^{i(kx-\om t)}$. As before, by transitioning to slow variables, we easily obtain the system:
\begin{multline}\label{system3}
    \big(\om-k\big)A[\varepsilon n] + B[\varepsilon n] + B[\varepsilon(n-1)] +\gamma C[\varepsilon n] + 
   \gamma C[\varepsilon(n-1)]+\mu_a \big(A[\varepsilon n]\big)^3=0,
    \\   
  \big(\om+k\big)B[\varepsilon n] + A[\varepsilon n] + A[\varepsilon(n+1)] +\mu_b \big(B[\varepsilon n]\big)^3=0
    \\
   \big(\om+k\big)C[\varepsilon n] + \gamma A[\varepsilon n] + \gamma A[\varepsilon(n+1)] +\mu_c \big(C[\varepsilon n]\big)^3=0.
    \\
\end{multline}
We will look for the solution to this system, as before, in the form of a series, for example, for $A$:
\begin{equation}
    A[\varepsilon n] = \varepsilon a_1 + \varepsilon^2 a_2 + \varepsilon^3 a_3+...
\end{equation}
and as a consequence
\begin{equation}
    A[\varepsilon (n-1)] = \varepsilon a_1 + \varepsilon^2 \big(a_2- \frac{\prt a_1}{\prt x}\big) + \varepsilon^3 \big( a_3 -\frac{\prt a_2}{\prt x} + \frac{1}{2}\frac{\prt^2 a_1}{\prt^2 x^2}\big)+...
\end{equation}
Substituting these expressions into system \eqref{system3} leads to the following relationships at different degrees of $\varepsilon$.

$\varepsilon:$
\begin{equation}
c_1 = \gamma b_1, \quad b_1 = -\frac{2a_1}{\om+k_0}
\end{equation}

$\varepsilon^3:$
\begin{equation}
\ddot{a}_1 = a_1-2\theta (a_1)^3,
\end{equation}
where
\begin{equation}\label{T3}
\theta = -\frac{1}{(1+\gamma^2)}\bigg[ (\om + k_0)\mu_a + \frac{16}{(\om + k_0)^3}\big(\mu_b + \gamma^4 \mu_c \big) \bigg].
\end{equation}
where we have a soliton only if $\theta>0$, which imposes certain restrictions on the nonlinear coefficients. Obviously, solitons can be obtained in different configurations of focusing and defocusing waveguides.
\begin{figure}
    \centering
    \includegraphics[scale = 0.8]{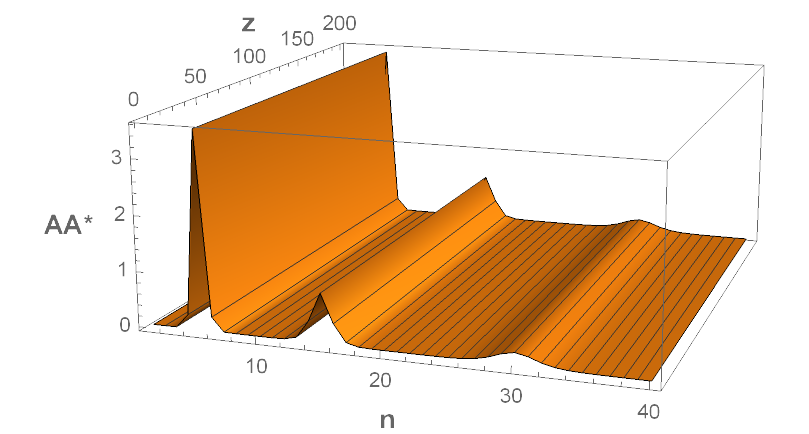}
    \caption{Evolution of intensity of three discrete solitons in the system with parameters $\gamma = 1$, $\om = 3$, $\mu_a = \mu_b=\mu_c = -1$. Parameters of $\varepsilon$ for solitons from left to right: $2,1,0.5$.}
    \label{fig3}
\end{figure}

\section{Conclusions}

We examined an array of waveguides consisting of three types of waveguides. We studied solitons in such a system as envelopes of wave packets modulated exponentially by amplitude. It was shown that it is possible for such waves to propagate in all waveguides simultaneously (see the second section), as well as in a discrete soliton that includes multiple waveguides.

The search for solutions was carried out within the framework of perturbation theory, or as it is often called, the theory of multiple scales. Despite the fact that the amplitudes of all solitons in this approach are assumed to be of the order of a small perturbation coefficient $\varepsilon$, numerical calculations showed that for the discrete soliton $\varepsilon\ll 1$ is not really a constraint, as seen in Figure \ref{fig3}.

Soliton solutions can be obtained with various combinations of focusing and defocusing waveguides, the main thing is $\theta > 0$ in equations \eqref{T1},\eqref{T2},\eqref{T3}.

\section{Acknowledgment}
I would like to express my gratitude to my teacher A.M. Kamchatnov for his constant help in my work. This method for finding soliton solutions was suggested to me by him.

\section{ Sources of funding}
The work was carried out as part of the state assignment dated 20.03.2025 No. 103-00001-25-02 on the topic
”Optical properties of dimer chains of laser waveguides
with topological defects, including nonlinear ones and
those with a negative refractive index.”

\end{document}